\documentclass{article}
\usepackage[accepted]{vietnam} 
\usepackage{natbib}
\usepackage{graphicx}      
\begin{document}
\twocolumn[
\title{Variability of Deeply Embedded Protostars: A New Direction for Star Formation?}
\titlerunning{Variability of Deeply Embedded Protostars}
\author{Doug Johnstone}{Doug.Johnstone@nrc-cnrc.gc.ca}
\address{NRC Canada, Herzberg Astronomy and Astrophysics, 5071 West Saanich Rd, Victoria, BC, Canada}

\vskip 0.5cm 
]

\begin{abstract}
The formation of a star is a dynamic process fed by the gravitational collapse of a molecular cloud core. Theoretical models and observations suggest that the 
majority of this infalling material settles into a protoplanetary disk before reaching the (proto)star and therefore that disk accretion processes are responsible for the rate
at which the (proto)star grows. There is no fundamental reason why infall and disk accretion need to be instantaneously identical. Indeed, even within the disk it might be anticipated that there are regions of strong and weak accretion. Together these facts suggest that (proto)stellar mass assembly should be both secular and stochastic and that the underlying physical processes leading to these time-variable accretion rates should manifest in observable time-dependent accretion luminosity variations.
\end{abstract}

\section{Introduction}
Observations of the temporal variability of stars have provided significant physical insight into stellar structure, e.g. internal structure and convection.  Recently pulsating pre-main sequence stars have also been identified and compared against star formation and evolutionary models \citep{zwintz08}. Given that deeply embedded protostars are primarily powered by mass infall from the surrounding protoplanetary disk, and that this accretion mass is anticipated to vary in both a secular and stochastic manner, long term monitoring of very young protostars should provide significant physical insight into their evolution and the underlying processes responsible for stellar assembly.

\section{Background}

The simplest models for the formation of a protostar begin with an isothermal cloud core, perhaps truncated by a bounding pressure \citep{bonnor56, ebert55}, collapsing from the inside-out. The steady-state infall accretion rate, from the cloud core, is thus explicitly determined by the physical parameters and in the case of the singular isothermal sphere reduces to $\dot M \sim c_s^3/G$, where $c_s$ is the sound speed in the core \citep{shu77}. For typical core conditions, this leads to $\dot M \sim 10^{-6}\,$M$_\odot\,$yr$^{-1}$. More sophisticated isothermal models typically retain the same order of magnitude for the accretion luminosity but include an initial burst of infall and a later decline \citep[e.g.][]{foster93}. Even the addition of angular momentum and magnetic fields does not significantly affect the cloud infall rate, although the material is often no longer able to stream directly to the central forming protostar \citep[e.g.][]{terebey}. Adding confidence to these fiducial calculations, the typical time over which a star accretes most of its mass has been measured to be $\sim0.5\,$Myrs \citep{evans09}, in reasonable agreement with the above steady-state infall rates.

The predicted accretion luminosity from these models, assuming that the infalling core material makes its way directly to the protostar surface, is significantly higher than that which is observed for the majority of protostars \citep{dunham10, hartmann}. To explain this dichotomy, various episodic accretion processes have been suggested in the literature in order to maintain the time-averaged accretion rate required for the growth of a (proto)star while also yielding a lower accretion rate (and luminosity) over much of the protostellar lifetime. Suggestions for episodic mechanisms include competitive accretion \citep[e.g.][]{offner11} and protostellar disk instabilities \citep[e.g.][]{zhu10, vorobyov05}. 

At the 2016 ``Star Formation in Different Environments" ICSE conference in Quy Nhon, Vietnam, for which this conference proceeding is a record, a wide variety of talks and posters presented both theoretical and observational evidence for episodic accretion and variability during the protostellar phase. Theoretically, \citet{kuffmeier} presented evidence for both secular and stochastic accretion onto protostellar sink cells created within large simulations of turbulent molecular clouds. On a smaller scale, \citet{lomax} found that only through episodic accretion and applying specific initial kinematic conditions within the originating core could protostars  form in their simulations {\it and} match the known observations. Observationally, \citet{tokuda} presented ALMA observations of complex velocity structure within the L1521F core, suggesting that the initial conditions of star formation can be much more dynamical than the simple isothermal models discussed above. As well, \citet{tuananh} determined that the envelope infall rate 140 AU from L527 was a factor of two different than the luminosity-derived accretion rate onto the central star. Considering the evolution of an ensemble of low-mass protostars, \citet{takakuwa} showed that large disks are formed very early during protostellar evolution and that when these disks are observable they already show evidence for structure, such as rings and gaps. Theoretical models for steady accretion through disks do not predict this intricate internal structure. On a different tack, \citet{stecklum} presented time-dependent observations of light-echoes from high-mass young stellar objects, showing specifically the eruptive behaviour of HMYSO S255IR-NIRS3.

\section{Observing Strategy} 

All of these theoretical and observational results support the notion that time variability of protostars is worth further investigation. Furthermore, while the large amplitude accretion events such as the observed FUors and EXors \citep[e.g.][]{herbig77, hartmann96} and the theoretically postulated gravitational instability in  disks \citep[e.g.][]{vorobyov05} may be extremely infrequent, shorter term monitoring provides clues to the accretion processes working in the inner protoplanetary disk where planet formation is expected to take place. To date, only three outbursts have been detected from deeply embedded protostars \citep{fisher,safron,hunter}, although indirect evidence for episodic accretion are found in both the shock-bullets seen in protostellar outflows \citep[e.g.][]{reipurth,raga} and non-equilibrium chemistry \citep{kim, jorgensen}.

Variability surveys of embedded protostars need to be performed in the far infrared through the sub-millimetre since the photons formed in the boundary layer at the stellar surface are reprocessed through the protostellar envelope \citep[see for example,][]{johnstone13}. Semi-analytic calculations show that the size of the inner, optically thick, portion of the protostellar envelope sets an effective lower limit to the speed with which an instantaneous change in source brightness will be visible to a distant observer, effectively limiting our ability to observe luminosity variations on timescales shorter than a few days to a week \citep{johnstone13}. For longer timescales, months to years, there exists little to no fundamental understanding of the amplitude of variability that might come from episodic accretion onto the embedded protostars. Observations are therefore essential.

The optimum wavelength to observe variability in deeply embedded protostars is the far infrared, where the spectral energy distribution peaks. Without an on-going far infrared space mission, however, observations at longer, sub-millimetre wavelengths take centre stage. An international team of astronomers, headed by myself and Greg Herczeg (Kavli, Peking), have obtained time at the James Clerk Maxwell Telescope (JCMT) to survey eight nearby ($d < 500$\,pc) star-forming regions with a 30 arc-minute field of view, at a monthly cadence over three years using the continuum camera SCUBA-2 at 850\,$\mu$m \citep{herczeg17}. Table 1 provides details on the eight regions, including the location, the number of bright sub-millimetre peaks, and a census of known protostars.  Figure 1 shows a sample field observed with the JCMT, overlaid with the locations of known protostars and disks.

This JCMT project is very much an exploratory survey and a stringent test of the ability to calibrate sub-millimetre observations. The relative calibration of the JCMT is measured to be $\sim 5$\% \citep{dempsey13} when using the standard data reduction pipeline. We have investigated a variety of techniques for performing `self-calibration' of our observations in order to improve on this value and have settled on a Gaussian clump-fitting analysis both to remove the pointing error of the telescope (leaving less than 1$''$ residual pointing error as compared to the 15$''$ beam) and to flux calibrate the fields to better than 3\% \citep{mairs17}. With one year of observations complete, a data reduction pipeline in place, and a team of dedicated astronomers considering each region carefully, a paper detailing the initial results will be written soon. Furthermore, we are also re-reducing earlier JCMT observations of these same star-forming regions, taken as part of the JCMT Gould Belt Survey \citep{wardthompson}, to look for secular variations over multi-year baselines. 

\section{Conclusions}

The mass accretion rate onto forming protostars is very likely to be time-dependent and thus observable through changes in the accretion luminosity of deeply embedded protostars. While there is strong evidence, both theoretical and observational, for large amplitude variations over extended periods of time, little is
known about the variability of protostars on shorter timescales. Measuring the amplitude and time variability over month to several year timescales will provide a
very powerful input and constraint for mass assembly and viscous accretion models, probing conditions within several AU of the forming protostar.

\begin{figure*}
\vskip 0.cm
\centering
$\begin{array}{cc}
\includegraphics[angle=0,width=14.cm]{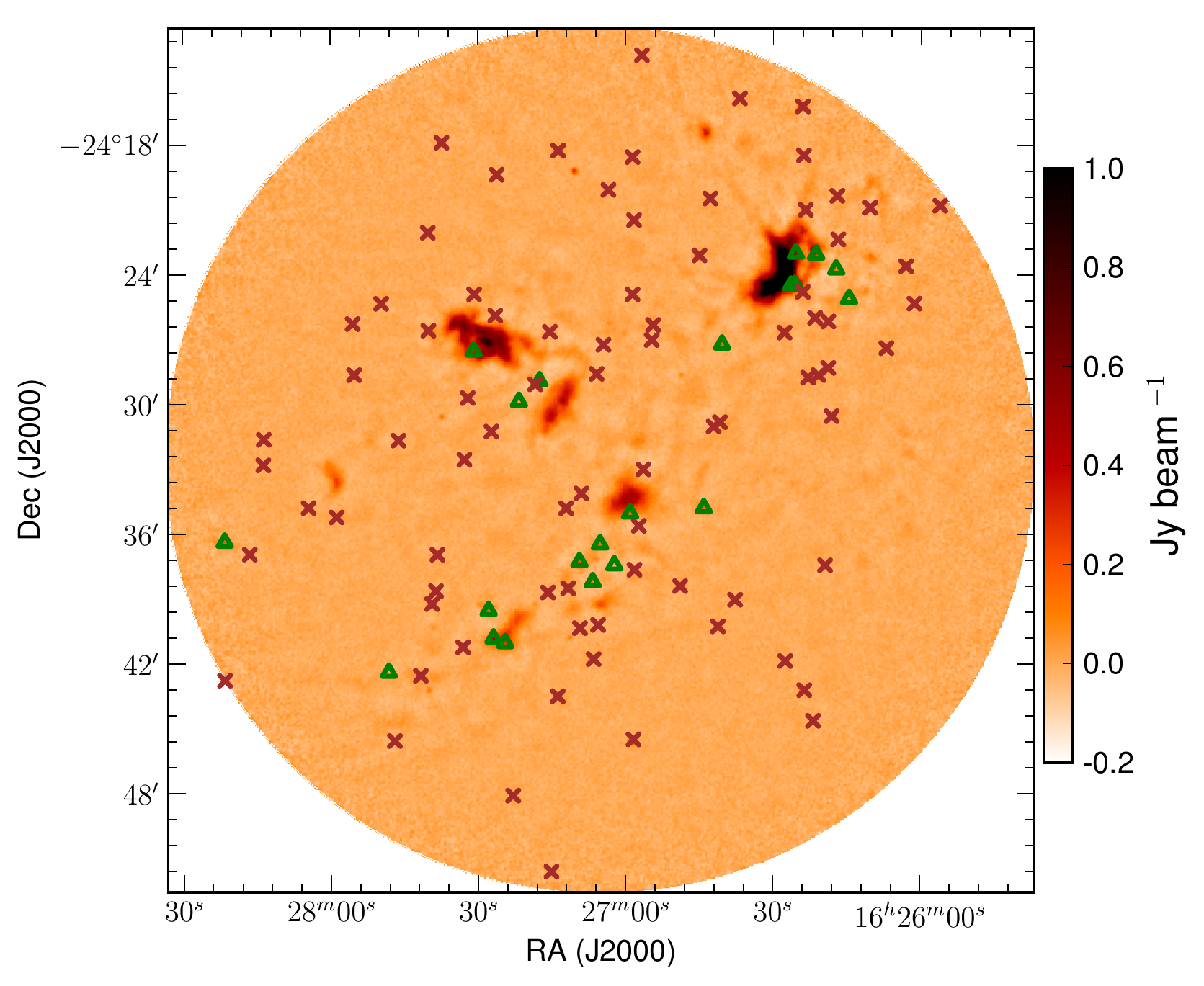} 
\end{array}$
\caption{ Ophiuchus Field. Image shows 850\,$\mu$m dust continuum emission. The location of YSOs are overlaid. Green triangles denote deeply embedded protostars while red crosses denote Class II disk sources.}
\label{fig:4}
\vspace{-0.5cm}
\end{figure*}

\begin{table*}
\begin{center}
\caption{JCMT Transient Survey Fields}
\medskip
\begin{tabular}{|ll|ccc|ccc|}
\hline
& & \multicolumn{3}{c}{SCUBA-2 peak flux/beam}& \multicolumn{3}{|c|}{Spitzer Sources}\\
Name&Location&{$> 0.2$ Jy}&{$> 0.5$ Jy}&{$> 1.0$ Jy}&{Class 0/I}&{Flat}&{Class II}\\
\hline
Perseus - NGC1333& 032854+311652& 27& 9& 5& 31& 13& 57\\
Perseus - IC348& 034418+320459& 5& 3& 2& 11& 6& 94\\
Orion A - OMC2/3& 053531-050038& 60& 30& 17& 32& 29& 158\\
Orion B - NGC2024& 054141-015351& 21& 9& 5& 11& 12& 87\\
Orion B - NGC2071& 054613-000605& 25& 11& 4& 14& 5& 54\\
Ophiuchus&162705-243237& 26& 5& 2& 18& 27& 60\\
Serpens Main& 182949+011520& 14& 10& 7& 18& 10& 47\\
Serpens South& 183002-020248& 20& 5& 2& 47& 30& 113\\
\hline
\end{tabular}
\end{center}
\end{table*}

\section*{Acknowledgments}
\vspace{-0.3cm}
This work was in part supported by an NSERC Discovery Grant.



\begin{thebibliography}
\bibitem[Bonnor(1956)]{bonnor56} Bonnor, W.B.\ 1956, MNRAS, 285, 201
\bibitem[Dempsey et al.(2013)]{dempsey13} Dempsey, J.T., et al.\ 2013, MNRAS, 430, 2534
\bibitem[Dunham et al.(2010)]{dunham10} Duhman, M.M, Evans, N.J.II, Terebey, S. Dullemond, C.P., \& Young, C.H.\ 2010, ApJ, 710, 470
\bibitem[Ebert(1955)]{ebert55} Ebert, R.\ 1955, ZA, 37, 217
\bibitem[Evans et al.(2009)]{evans09} Evans, N.J.II, Dunham, M.M, J{\o}rgensen, J.K., et al.\ 2009, ApJS, 181, 321
\bibitem[Fisher et al.(2012)]{fisher} Fisher, W., et al. 2012, ApJ, 756, 99
\bibitem[Foster \& Chevalier(1993)]{foster93} Foster, P.N. \& Chevalier, R.A.\ 1993, ApJ, 416, 303
\bibitem[Hartmann et al.(2016)]{hartmann} Hartmann, L., Herczeg, G., \& Calvet, N.\ 2016, ARAA, 54, 135
\bibitem[Hartmann \& Kenyon(1996)]{hartmann96} Hartmann, L.\ \& Kenyon, S.J.\ 1996, ARA\&A, 34, 207
\bibitem[Herczeg et al.(2017)]{herczeg17} Herczeg, G.\ et al. in preparation
\bibitem[Herbig(1977)]{herbig77} Herbig, G.H. 1977, ApJ, 217, 693
\bibitem[Hunter et al.(2017)]{hunter} Hunter, T.R., Brogan, C.L., MacLeod, G., et al.\ 2017, APJL, accepted
\bibitem[Johnstone et al.(2013)]{johnstone13} Johnstone, D., Hendricks, B., Herczeg, G.J., \& Bruderer, S. 2013, ApJ, 765, 133
\bibitem[J{\o}rgensen et al.(2013)]{jorgensen} J{\o}rgensen, J., et al.\ 2013, ApJ, 779, L22
\bibitem[Kim et al.(2011)]{kim} Kim, H.J., et al.\ 2011, ApJ, 729, 84
\bibitem[K\"uffmeier et al.(2017)]{kuffmeier} K{\"u}ffmeier, M., Haugb{\o}lle, T., \& Nordlund, A.\ 2017, this volume
\bibitem[Lomax et al.(2017)]{lomax} Lomax, O., Whitworth, A.P., \& Hubber, D.A.\ 2017, this volume
\bibitem[Mairs et al.(2017)]{mairs17} Mairs, S., et al.\ 2017 in preparation
\bibitem[Offner \& McKee(2011)]{offner11} Offner, S.S.R.\  \& McKee, C.F.\  2011, ApJ, 736, 53
\bibitem[Raga et al.(2002)]{raga} Raga, A.C., et al. 2002, A\&A, 392, 267
\bibitem[Reipurth(1989)]{reipurth} Reipurth, B.\ 1989, Nature, 340, 42
\bibitem[Safron et al.(2015)]{safron} Safron, E.J., et al.\ 2015, ApJL, 800, 5
\bibitem[Shu(1977)]{shu77} Shu, F.H.\ 1977, ApJ, 214, 488
\bibitem[Stecklum et al.(2017)]{stecklum} Stecklum, B., Heese, S., Wolk, S., Caratti o Garatti, A., Ibanez, J.M., \& Linz, H.\ 2017, this volume
\bibitem[Takakuwa et al.(2017)]{takakuwa} Takakuwa, S., et al.\ 2017, this volume
\bibitem[Terebey, Shu, \& Cassen(1984)]{terebey} Terebey, S., Shu, F.H., \& Cassen, P.\ 1984, ApJ, 286, 529
\bibitem[Tokuda(2017)]{tokuda} Tokuda, K.\ 2017, this volume
\bibitem[Tuan-Anh et al.(2017)]{tuananh} Tuan-Anh, P., Nhung, P.T., Diep, P.N., Hoai, D.T., Phuong, N.T., Thao, N.T., \& Darriulat, P.\ 2017, this volume
\bibitem[Vorobyov \& Basu(2005)]{vorobyov05} Vorobyov, E.I.\ \& Basu, S.\ 2005, ApJ, 633, 137
\bibitem[Ward-Thompson et al.(2007)]{wardthompson} Ward-Thompson, D., et al.\ 2007, PASP, 858, 855
\bibitem[Zhu et al.(2010)]{zhu10} Zhu, Z., Hartmann, L., Gammie, C.F., et al.\ 2010 ApJ, 713, 1134
\bibitem[Zwintz(2008)]{zwintz08} Zwintz, K.\ 2008, ApJ, 673, 1088
\end{thebibliography}

\end{document}